# Optomechanical crystals for spatial sensing of submicron sized analytes

D. Navarro-Urrios[1,2*], E. Kang[3], P. Xiao[2], M. F. Colombano[1,2], G. Arregui[2], B. Graczykowski[3,4] N. E. Capuj[5,6], M. Sledzinska[2], C. M. Sotomayor-Torres[2,7], G. Fytas[3]

[1] MIND-IN2UB, Departament d'Enginyeria Electrònica i Biomèdica, Facultat de Física, Universitat de Barcelona, Martí i Franquès 1, 08028 Barcelona, Spain

[2] Catalan Institute of Nanoscience and Nanotechnology (ICN2), CSIC and BIST, Campus UAB, Bellaterra, 08193 Barcelona, Spain

[3] Max Planck Institute for Polymer Research, Ackermannweg 10, 55128 Mainz, Germany

[4] Faculty of Physics, Adam Mickiewicz University, Uniwersytetu Poznanskiego 2, 61614 Poznan, Poland

[5] Depto. Física, Universidad de La Laguna, 38200 San Cristóbal de La Laguna, Spain

[6] Instituto Universitario de Materiales y Nanotecnología, Universidad de La Laguna, 38071 Santa Cruz de Tenerife, Spain

[7] Catalan Institute for Research and Advances Studies ICREA, 08010 Barcelona, Spain

**Abstract**

Optomechanical crystal cavities have rich perspectives for detecting and indirectly analysing biological particles, such as proteins, bacteria and viruses. In this work we demonstrate the working principle of an optomechanical crystal cavity operating under ambient conditions as a sensor of submicrometer analytes by optically monitoring the frequency shift of thermally activated mechanical modes. The resonator has been specifically designed so that the cavity region supports a particular family of low modal-volume mechanical modes, commonly known as -pinch modes-. These involve the oscillation of only a couple of adjacent cavity cells that are relatively insensitive to perturbations in other parts of the resonator. The eigenfrequency of these modes decreases as the deformation is localized closer to the centre of the resonator. Thus, by identifying specific modes that undergo a frequency shift that amply exceeds the mechanical linewidth, it is possible to infer if there are particles deposited on the resonator, how many are there and their approximate position within the cavity region.

**Introduction**

In the last decade photonic crystals (PhCs) have been used as optical sensing platform, in particular for label-free biosensing [1,2]. PhCs cavities strongly confine light in ultra-low volumes, enabling the detection of chemical species down to nanometric dimensions when placed in close proximity to the cavity region by monitoring the spectral shift of an optical resonance. [3] Alternatively, sensors based on nano-electro-mechanical system (NEMS) can be extremely sensitive as mass and/or force sensors based on the frequency shift of the supported mechanical modes and/or variations of their quality factors. [4] Indeed, they can achieve mass resolution in the nanogram scale and resolve forces as small as 10 pN even when operating in a fluid environment. [5] Both sorts of devices, if combined with advanced chemical surface functionalization techniques and the integration in microfluidic systems, [6] pave the way towards ultra-compact lab-on-chip platforms with very high performance and real time monitoring. [7,8]

More recently, cavity optomechanical systems [9] have also been proposed as successful sensing platforms since they exhibit similar characteristics to those of NEMS while adding the possibility of optically transducing the mechanical motion even in liquid environments. [10] Mechanical modes can be thermally activated by Langevin forces or, by exploiting radiation pressure forces, driven to a regime of coherent high amplitude oscillations. [11] Indeed, in the latter regime, recent experiments have elegantly demonstrated an enhanced sensing resolution down to a single molecule, achieved by monitoring the change of the mechanical spring constant as the optical detuning between the laser and the resonance is modified. [12]

A particularly suited class of cavity optomechanical system for sensing applications are optomechanical crystal cavities (OMC's), nanostructured materials which simultaneously behave as photonic and phononic crystal cavities. [13] They can provide a combination of the sensing characteristics of PhCs and NEMS and extend further the sensing capabilities of these systems by exploiting optomechanical coupling effects, which are engineered by structural design. [14,15,16] In this work we present a novel OMC-based sensing device that addresses the issue of counting and spatially localizing submicrometric analytes. In fact, this is a pervasive issue that often compromises the analysis of the registered data. For instance, when dealing with standard NEMS sensors such as cantilevers or strings over which one or several analytes are leaning, their effect on the mechanical mode of the resonator significantly depends on their specific location.[17]

Here the proposed design exploits low modal-volume mechanical modes –pinch modes- that involve the oscillation of a few adjacent crossbars of the cavity cells and are relatively insensitive to perturbations in other parts of the resonator. We show that, in most cases, the mechanical modes are spatially modified in a way that the analyte blocks the affected crossbars due to its additional mass. Thus, the observed frequency shift is independent of the elastic properties of the analyte or its adhesion force to the OMC. In order to extract more information about the analyte, we discuss the possibility of bringing together its natural frequencies to that of the affected pinch mode so that both modes hybridize. We also consider the emergence of new mechanical modes involving the collective oscillation of the affected crossbars and the analyte, the frequency of which depends mostly on the mass of the analyte.

**Description of the geometries and experimental setup**

The fabricated structures are 1-dimensional silicon OMCs based on a unit cell consisting of a crystalline silicon rectangular block of lattice constant *a* with a rectangular hole. In this work we focus on the lowest energy frequency even-even mechanical band that is commonly known as 'pinch' mode band, which, for the present fabricated structure, lies below 0.6 GHz.[13,18] The band diagram has been computed with a finite elements method (FEM) solver both using an imported cell of the fabricated structure and the nominal cell (central panel of Figure 1a and Supplementary Information, respectively). The pinch mode is a localized, in-plane mechanical vibration and its deformation profile for a wave number $k_x=\pi/a$ in the axial direction of the OMC, i.e., at the X-point, is illustrated in Figure 1a.

The cavity region is highlighted in Figure 1d with a dashed yellow box and consists of an odd number of holes with the spacing between them reduced quadratically from the nominal lattice constant at the beam perimeter (*a*) to $\Gamma a$ for the cell in the centre of the OMC ($\Gamma<1$). On both sides of the OMC the nominal cell is repeated 10 times, thus acting as an effective mirror for optical and mechanical cavity modes. Optical modes are drawn up in energy from the X-point of a TE-like band, which is placed at about 170 THz. In this sense, the depth of the cavity region has been engineered so that it supports optical modes within the spectral range of our light sources, i.e., between 180 and 210 THz. Given that the pinch mechanical modes are drawn from a band edge at the X-point, each crossbar vibrates 180º out of phase with respect to its nearest neighbors. Consequently, vibration of neighboring cells add up in anti-phase to the overall optomechanical coupling. If the cells composing the cavity region are sufficiently different, a pinch cavity mode is localized just in one or two cells, otherwise it involves the deformation of several cells. In terms of the sensor design, this implies that the cavity region cannot be distributed over an arbitrary large number of cells, since the spatial resolution would be lost at the expense of increasing the sensing area. Thus, a compromise has been found considering the number of cells, pinch modes spatial localization and spectral resolution (>1 MHz, the experimental mechanical linewidth). The cavity region was therefore distributed over 27 cells and the maximum reduction of the pitch in the centre of the OMC was $\Gamma=0.9$. The total number of cells composing the OMC is 47 and the sensing area is about 12 $\mu m^2$ (see Figure 1d). As illustrated in Figure 1a by plotting the band edge at the X-point as a function of $\Gamma$, the confinement potential pushes the pinch band to lower frequencies. This implies that the pinch modes within the cavity region own eigenfrequencies ($\Omega_{m,o}$) of decreasing values as the modes are localized in cells closer to the centre. More details concerning the nominal design and the fabrication of the OMCs can be found in the Supplementary Information (Sections S1 to S3).

Submicrometer silica particles (diameter $d = 495 \pm 16$ nm) were dispersed in ethanol up to a concentration of 38 mg/L. After ethanol evaporation, several particles were deposited on the OMCs. Other particles remained underneath the OMCs leaning on the silicon substrate since the last ethanol volume prior to evaporation lies under the OMC. We ensured that these dragged particles were not in contact with the OMC by SEM inspection in a tilted configuration.

The experiments were made in a standard set-up to characterize optical and mechanical properties of OMCs (Figure 1). A tuneable infrared laser (L1) covering the spectral range 1440-1640 nm was connected to a tapered fiber in the shape of a microloop.[19] The polarization state of the light entering the tapered region was set with a polarization controller (FPC). The fibre was brought into contact with the etched frame, while the thinnest part of the fibre rests above the central region of the OM crystal. The long tail of the evanescent field, which is several hundreds of nanometers, and the low spatial resolution (~5 $\mu m^2$) of the tapered fibre allowed the local excitation of resonant optical modes of the OMC (Figure 1b).

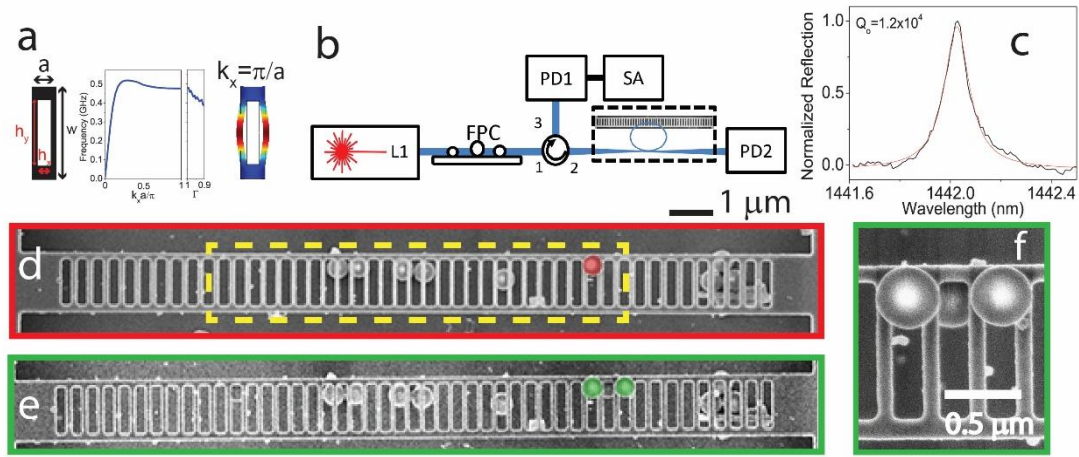

*Figure 1. Optomechanical crystal geometry and experimental setup. a) Left. Sketch of the unitary cell of the OMC. The nominal geometrical parameters are $\Lambda$=362 nm, w=1396 nm, $h_y$=992 nm, $h_x$=190 nm and thickness=250 nm. Centre. Band diagram of the pinch mechanical band and dependence of the X-point band edge energy on the reduction factor. Right. Deformation profile of the pinch mode at the X-point calculated with a FEM solver. The simulations have been performed by importing single mirror cells of the fabricated geometry and applying Floquet periodic conditions. b) Experimental setup used for the optomechanical characterization of the fabricated OMCs. c) Reflection spectra around the first optical resonance of the OMC. d) and e) SEM micrographs of the characterized OMCs with one and two submicrometer particles on top and have been highlighted in red and green, respectively. The cavity region is delineated by a dashed yellow box in d). f) Zoom of the region of the OMC holding the particles. The diameter of the particles is 495 ± 16 nm.*

The OMC is a bi-directional photonic cavity, hence it can optically couple to both forward and backward propagating fiber modes. By using a fiber circulator it is possible to collect both the reflected and transmitted signals, which are detected by InGaAs fast photoreceivers PD1 and PD2, respectively. Figure 1c shows a normalized reflected spectrum of the OMC under study around the highest energy optical resonance within the spectral range of the tuneable laser. The optical quality factor of this particular mode is on the order of $10^4$, while the coupled power fraction exceeds 90%. To check for the presence of a radiofrequency (RF) modulation of the reflected signal we connect the output of PD1 to the 50 Ohm input impedance of a signal analyser (SA) with a bandwidth of 13.5 GHz. All the measurements were performed in an anti-vibration cage at ambient conditions. The experimental approach requires the optical transduction of the mechanical spectrum of a specific OMC when there is one particle (Figure 1d) and two particles (Figures 1e and 1f) on the OMC. Both are compared to the reference mechanical spectrum of the OMC without particles.

The followed procedure to position the particles was first to deposit two particles on the OMC, which were leaning only two cells away from each other on the outer cells of the cavity region after the ethanol evaporated (Figure 1e). We then proceeded to remove the particle placed further from the centre (Figure 1d) and finally the remaining one using the tip of a tapered fibre. It is worth noting that the optical spectra did not suffer significant changes when depositing the particles. Indeed, the optical mode used in this study appeared at roughly the same spectral position with similar quality factors as that of Figure 1c during the whole experiment.

**FEM simulations of the fabricated OMC**

To model the fabricated OMC and account for the differences from the nominal design, the in-plane geometry was imported from the SEM micrographs into a FEM solver, where the thickness is that of the top Si layer of the SOI wafer. This procedure ensures a reasonably good agreement between the measured optical and mechanical modes and those extracted from simulations. We have also considered the case of including a spherical $SiO_2$ particle leaning on one of the crossbars of the OMC, aiming to mimic the situation reported in the SEM micrograph of Figure 1d.

Figure 2b shows the computed spectral dependence of the single-particle OM coupling rate ($g_o/2\pi$) without the particle taking into account photo-elastic and moving-boundary contributions (see SI, Sections S2 and S3). [19] The optical mode used for these calculations (see Figure 2a) appears at around 200 THz and it was checked that it is not meaningfully affected by the presence of the particle (not shown), which is consistent with the experiment observation. We measure a sizeable reduction of the expected values of $\Omega_{m,o}$ with respect to the nominal design (details in SI, Section S2) associated to an effective increase of $h_x$ of the fabricated geometries. In order to mimic the experimental RF spectrum, we have represented each mode by a Lorentzian line shape and added up the contributions of all modes. The inspection of the spatial profile of the mechanical modes providing large $g_o$ supports their association to pinch-like modes, also confirming that $\Omega_{m,o}$ decreases as the modes become localized closer to the centre of the OMC (details in SI, Section S3). Notably, not all pinch modes display large coupling $g_o$ values, so these modes would be eventually hidden below the noise level of the experiment. The mass and modal volume of such modes are on the order of 0.1 pg and 0.1 $\mu m^3$, respectively.

Figure 2c reports the frequency shift of the mechanical modes ($\Delta\Omega$) with respect to the original eigenfrequency $\Omega_{m,o}$ when the particle is introduced. We observe a significant shift towards higher frequencies exceeding 0.5% whenever the original mode involves the crossbar with the particle. As expected, the modes affected to a greater extent appear in the highest frequency range, given that the particle is placed almost at the boundary of the cavity region and higher frequency modes tend to localize away from the center (details in SI, Section S3). This is illustrated for the mode highlighted with the green dashed box, which displays both large coupling $g_o$ and $\Delta\Omega$ values. The deformation profiles of that particular mode with and without the particle are displayed in Figure 2d (bottom and top panels, respectively). Since the presence of the particle substantially increases the overall mass of the crossbar, the original mode accommodates to a new spatial configuration in which that crossbar remains fixed. We have also modified the Young modulus and density of the particle and its position along the crossbar and verified that the modified mode does not change further its frequency or spatial distribution (see SI, Section S5). The previous statement holds unless there are mechanical eigenfrequencies of the isolated particle that are similar to those of the original pinch modes, in which case the modes hybridize (see SI, Section S6) similarly to what was experimentally demonstrated in Ref. [20] by coupling the vibrating modes of a *S. epidermidis* bacterium to those of an optomechanical disk resonator. Interestingly, the vibrating modes of that particular bacterium have an expected fundamental frequency of $500 \pm 100$ MHz [20], which is compatible with the range covered by the pinch modes of the OMC reported here. Regarding other spectral regions of interest, mechanical modes involving the deformation of the silica particle (*d*=495 nm) start appearing at much higher frequencies (few GHz), consistent with those observed in an isolated particle of the same size (see SI, Sections S6 and S7). Another feature worth mentioning is the onset of pinch-like modes that involve the collective oscillation of the particle and the bar in contact with it (details in SI, Section S6). These latter modes appear at significantly lower frequencies than those shown in Figure 2, given that their mass is about a factor of two larger.

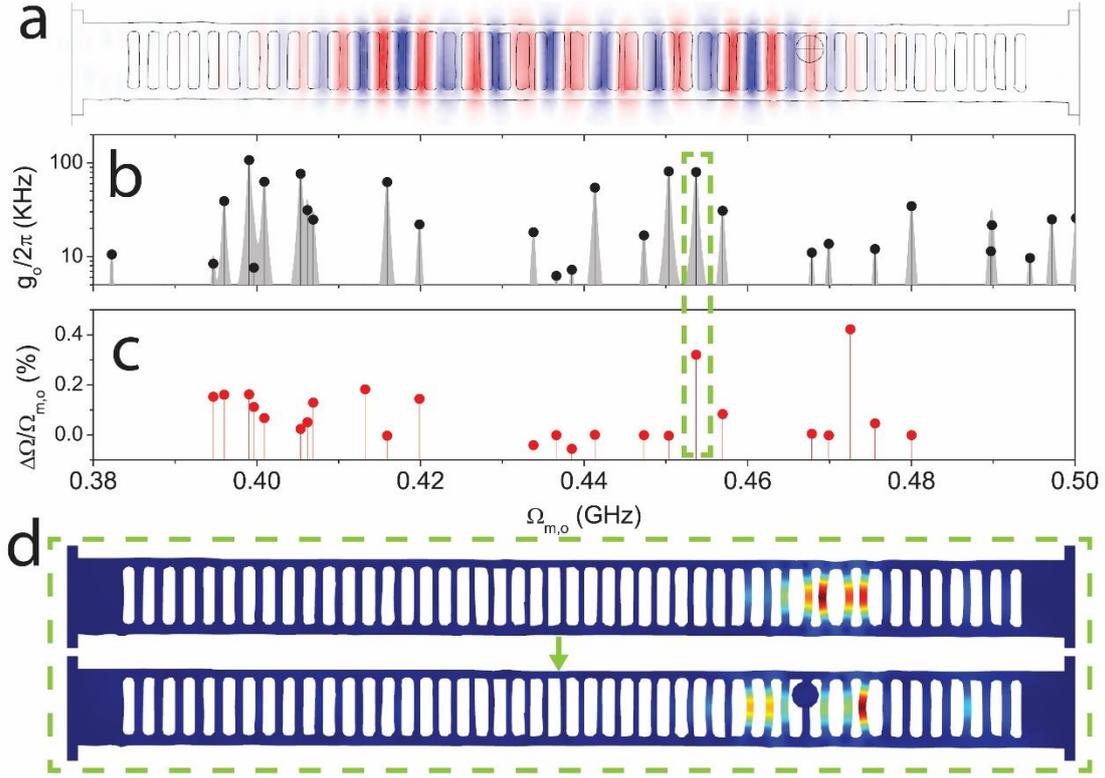

*Figure 2. Finite-Element-Method simulations of the fabricated OMC with and without a submicrometric particle. a) Spatial profile of the simulated optical mode employed for the evaluation of the single-particle optomechanical coupling rates ($g_o/2\pi$). b) Computed $g_o/2\pi$ covering the frequency range of the pinch mechanical modes family. c) Relative frequency shift of the mechanical modes with respect to the eigenfrequencies displayed by the OMC without particles. The dashed green box highlights the case of the mechanical modes visualized on panel d). The geometry has been imported from the SEM micrograph of the fabricated OMC.*

**Experimental results**

When light is coupled to the optical mode reported in Figure 1c, thermally driven motion of the mechanical modes is seen by processing the reflected light with the spectrum analyser. Here, mechanical modes with non-negligible $g_o$ appear as narrow Lorentzian peaks in the frequency spectrum, as reported in the bottom panel of Figure 3. In the specific region between 0.35 GHz and 0.55 GHz the OMC presents strong transduced signal associated to the pinch mechanical mode family. As expected, there is a rich substructure of peaks related to different cavity pinch modes coming from the same original mechanical band. Mechanical quality factors on the order of $10^2$ are measured in this frequency region, which translate into mechanical linewidths of about 1-2 MHz (less than 0.5% of $\Omega_{m,o}$). Given the frequency range of the modes, mechanical losses are probably dominated by viscoelastic losses due to interaction with the surrounding medium [21]. The differences between computed (Figure 2b) and experimental (black curve of Figure 3a) $\Omega_{m,o}$ are probably associated to a non-negligible deviation of the sidewalls of the OMCs from verticality, contrary to what is assumed in the simulations.

The black curve in Figure 3a displays the mechanical spectrum of the as-fabricated OMC. When compared with the curve associated to the OMC with one particle on it (red curve), it is possible to confirm that most peaks remain relatively unaltered up to 0.48 GHz. However, above that frequency and up to about 0.55 GHz (highlighted frequency range) there is one RF peak that is strongly affected, while the others remain almost unaltered. Only a couple of the remaining RF peaks are modified as well by adding the second particle on the OMC (green curve), while the RF peaks that changed already by the inclusion of the first particle are unaffected. For instance, the RF peak appearing slightly above 0.5 GHz is only sensitive to the presence of the first particle, while the adjacent one is only sensitive to the presence of the second particle. The modified pinch modes are placed in a spectral region that would be associated to mechanical vibrations on the sides of the cavity region, which is in good agreement with the position of the particles measured by SEM as illustrated in the simulations of Figure 2. These results demonstrate that this particular design of OMC enables the detection of single particles and the identification of the region of the resonator in which they are placed. To better illustrate the observations made on the RF spectra, in Figure 3b we represent the frequency shift of the modes ($\Delta\Omega$) with respect to the original eigenfrequency $\Omega_{m,o,}$ for the two cases under study, i.e., one particle (red curve) and two particles (green curve). We include the mechanical linewidth limit of about 0.5% as a horizontal dashed line. This level is clearly overcome by three of the peaks under analysis within the highlighted range, demonstrating that it is relatively straightforward to detect the frequency shift caused by the presence of the particles. The obtained values exceeded those extracted from the simulations, probably because the contact region has not been accurately reproduced. In fact, it seems that from the SEM images the nanoparticles contact the OMC in more than one place, affecting two crossbars instead of only one. In any case, a qualitative good agreement with the simulations has been found. Finally, when removing the last particle, the mechanical spectrum resembles the original one, without particles, with minor relative shifts below the mechanical linewidth of the resonances.

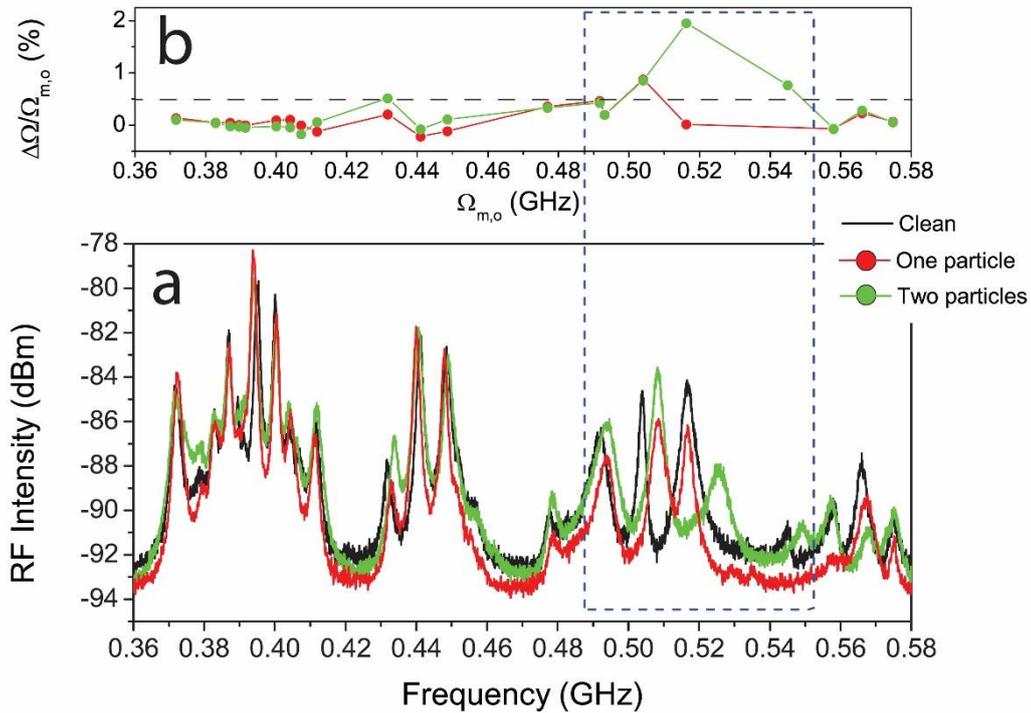

*Figure 3. Experimental demonstration of the sensing principle a) Transduced mechanical modes in the frequency range of the pinch modes family for the case of the as-fabricated OMC (black), the OMC with one particle (red) and with two particles (green). b) Relative frequency shift of the mechanical modes with respect to the eigenfrequencies displayed by the as-fabricated OMC. The horizontal black dashed line corresponds to the typical mechanical linewidth of the observed modes. The blue dashed box highlights the region in which the mechanical modes are significantly affected by the presence of particles.*

In summary, we have demonstrated the working principle of a novel sensor based on the modification, with respect to an as-fabricated reference, of the frequencies of the "pinch" mechanical modes of an optomechanical crystal cavity of the ladder type. By inspecting which of the original modes are affected, we have experimentally verified that the working principle of the sensor enables the determination of the position of submicrometer sized analytes of spherical shape along the cavity region and quantification of how many particles/analytes are deposited. These quite unique experimental and numerical observations also open the possibility of extracting a rough upper estimate of the dimensions of the analyte. The described technique, the accuracy of which can increase by performing a prior calibration of the OMC, reveals to be robust given that, in general, the particle blocks the contacted crossbar so that the modifications of the mechanical spectrum do not depend on the size or elastic properties of the analyte. As an exception, FEM simulations predict that if the mechanical eigenfrequencies of the original modes and those of the analyte are brought together to the same frequency, they are expected to hybridize, so that it would be possible to also extract information about the elastic properties of the analyte and even from the adhesion forces. The relatively standard fabrication requirements and large sensing response of the reported optomechanical sensor open the possibility of operating it in a fluidic environment and explore surface functionalization techniques for detection of specific biological targets such as such as proteins, bacteria and viruses, in particular the new described variant of coronavirus (COVID-19).

ACKNOWLEDGEMENTS

D. N. U. gratefully acknowledges the support of a Ramón y Cajal postdoctoral fellowship (RYC-2014-15392) and the Ministry of Science, Innovation and Universities (PGC2018-094490-B-C22). ICN2 is supported by the Severo Ochoa program from the Spanish Research Agency (AEI, grant no. SEV-2017-0706) by the CERCA Programme / Generalitat de Catalunya and by the Ministry of Science, Innovation and Universities (PGC2018-101743-B-I00). G. A and P.X acknowledge the support of a BIST and COFUND PREBIST studentships, respectively. E.K. and G.F. acknowledge the financial support by ERC AdG SmartPhon (Grant No. 694977).

# Supplementary Info: Optomechanical crystals for spatial sensing of submicron sized analytes

D. Navarro-Urrios[1,2*], E. Kang[3], P. Xiao[2], M. F. Colombano[1,2], G. Arregui[2], B. Graczykowski[3,4] N. E. Capuj[5,6], M. Sledzinska[2], C. M. Sotomayor-Torres[2,7], G. Fytas[3]

[1] MIND-IN2UB, Departament d'Enginyeria Electrònica i Biomèdica, Facultat de Física, Universitat de Barcelona, Martí i Franquès 1, 08028 Barcelona, Spain

[2] Catalan Institute of Nanoscience and Nanotechnology (ICN2), CSIC and BIST, Campus UAB, Bellaterra, 08193 Barcelona, Spain

[3] Max Planck Institute for Polymer Research, Ackermannweg 10, 55128 Mainz, Germany

[4] Faculty of Physics, Adam Mickiewicz University, Umultowska 85, 61614 Poznan, Poland

[5] Depto. Física, Universidad de La Laguna, 38200 San Cristóbal de La Laguna, Spain

[6] Instituto Universitario de Materiales y Nanotecnología, Universidad de La Laguna, 38071 Santa Cruz de Tenerife, Spain

[7] Catalan Institute for Research and Advances Studies ICREA, 08010 Barcelona, Spain

## S1. Band diagrams of the nominal unit cell

Figure 1S displays photonic and phononic band diagrams (left and right panels, respectively) corresponding to the nominal design, whose geometrical parameters have been written in terms of $a$ and $\Gamma$. Frequencies $\omega$ are given with dimensionless units, i.e., they are divided by $c/a$, where $c$ is either the velocity of light in vacuum for electromagnetic waves or the transverse speed of sound in silicon in [100] direction ($c_t$=5844 m/s). In the case of the photonic dispersion, we have only considered two relevant TE-polarized ($y$-direction) bands that define a TE photonic gap between $\omega$=0.2 and $\omega$=0.25. In the case of phononic dispersion, we have considered the lowest energy band displaying an even-even symmetric band with respect to the $xy$ and $xz$ planes. The modes that verify these conditions are the "pinch" mechanical modes represented in Figure 1 of the main text.

We have also studied the dependence of the energy of the Bloch modes at the x-point with $\Gamma$. Regarding the photonic bands, optical modes are drawn up in energy by decreasing $\Gamma$. Thus, if the cavity region between the mirrors is constructed so that the pitch is gradually reduced towards the center, cavity optical modes are expected to appear owning frequencies slightly higher than that of the edge of the lower band. On the contrary, the "pinch" band is pushed down by decreasing $\Gamma$, so that the cavity mechanical modes would have slightly lower energies than that of the edge of the band.

The geometrical parameters of the fabricated unit cell have been rescaled to the values reported in the main text so that the spectral range of the tunable laser (around 200 THz) covers the frequency region swept by the lower photonic band edge when $\Gamma$ is varied between 1 and 0.9.

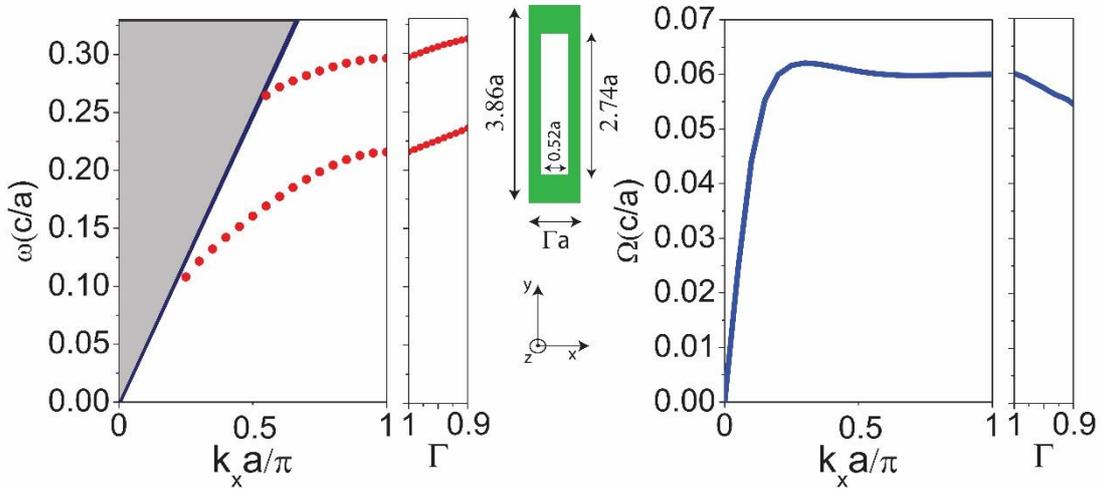

**Figure S1.** Photonic (left) and phononic (right) normalized band diagrams of the unit cell depicted in green. We have only considered the relevant bands for the purpose of this work, i.e., TE-polarized modes and "pinch" modes for the photonic and phononic calculations, respectively.

## S2. OM coupling calculations

Single-particle Optomechanical (OM) coupling rates ($g_O$) between optical and mechanical modes are estimated by taking into account both photo-elastic (PE) and moving-interface (MI) effects [1,2,3]. The PE effect is a result of the acoustic strain within bulk silicon while the MI mechanism comes from the dielectric permittivity variation at the boundaries associated with the deformation.

The calculation of the MI coupling coefficient $g_{MI}$ is performed using the integral given by Johnson et al. [1];

$$g_{MI} = -\frac{\pi\lambda_r}{c} \frac{\oint (\mathbf{Q}\cdot\hat{\mathbf{n}})(\Delta\varepsilon \mathbf{E}_{\parallel}^2 - \Delta\varepsilon^{-1}\mathbf{D}_{\perp}^2)dS}{\int \mathbf{E}\cdot\mathbf{D}\,dV} \sqrt{\hbar/2m_{eff}\Omega_m} \tag{S1}$$

where $\mathbf{Q}$ is the normalized displacement (max{|$\mathbf{Q}$|}=1), $\hat{\mathbf{n}}$ is the normal at the boundary (pointing outward), $\mathbf{E}$ is the electric field and $\mathbf{D}$ the electric displacement field. $\varepsilon$ is the dielectric permittivity, $\Delta\varepsilon = \varepsilon_{silicon} - \varepsilon_{air}$, $\Delta\varepsilon^{-1} = \varepsilon^{-1}_{silicon} - \varepsilon^{-1}_{air}$. $\lambda_r$ is the optical resonance wavelength, $c$ is the speed of light in vacuum, $\hbar$ is the reduced Planck constant, $m_{eff}$ is the effective mass of the mechanical mode and $\Omega_m$ is the mechanical mode eigenfrequency, so that $\sqrt{\hbar/2m_{eff}\Omega_m}$ is the zero-point motion of the resonator.

A similar result can be derived for the PE contribution [2,3]:

$$g_{PE} = -\frac{\pi\lambda_r}{c} \frac{\langle E|\delta\varepsilon|E\rangle}{\int \mathbf{E}\cdot\mathbf{D}\,dV} \sqrt{\hbar/2m_{eff}\Omega_m} \tag{S2}$$

where $\delta\varepsilon_{ij} = \varepsilon_{air} n^4 p_{ijkl} S_{kl}$, being $p_{ijkl}$ the PE tensor components, $n$ the refractive index of silicon, and $S_{kl}$ the strain tensor components.

The addition of both contributions results in the overall single-particle OM coupling rate:

$$g_O = g_{MI} + g_{PE} \tag{S3}$$

## S3. Design of the Optomechanical crystal

The defect used here consists of an odd number of cells, with the pitch reduced quadratically towards the center down to a value equal to $\Gamma a$. On both sides of the cavity region the nominal cell is repeated over 10 times, thus acting as an effective mirror for optical and mechanical cavity modes. FEM simulations have verified that adding more mirror cells does not improve further the quality factors of the cavity modes.

The cavity region has been distributed over a total number of cells denoted by $N$, which determines the sensing area ($S$). However, the defect region cannot be distributed over an arbitrary large number of cells. Indeed, if the various cells composing the defect region are not different enough among them, a pinch mode can involve the deformation of several cells. In terms of sensor design this implies that spatial resolution would be lost at the expense of increasing $S$. Thus, a compromise has been found in terms of having the largest number of cells composing the cavity region, i.e., the largest $S$ value, while also verifying that: i) pinch modes do not involve more than two cells and ii) the frequency difference between two adjacent modes is larger than the experimental mechanical linewidth (about 1 MHz). In Figure S2 we show the results of computing the normalized $g_O$ for different realizations of an OMC in which the number of cells composing the cavity region has been increased from $N=11$ to $N=31$. In order to resemble realistic RF spectra we have associated to each mode a Lorentzian distribution owning a linewidth equal to that required, finally adding up the contribution of all modes.

As expected, the pinch modes appear in the region between the bad edge values for $\Gamma=0.9$ and $\Gamma=1$ (left and right dashed areas, respectively) and their number roughly scales with $N$. The deformation profiles of the modes are localized more towards the centre as their frequency decrease, in agreement with what extracted from the study of the dependence of the band edge energy with $\Gamma$. Up to $N=27$ all the computed modes involve at most two cells and display a frequency separation that in average fulfils the requirements. This starts to be no longer valid for larger $N$ values. This can be observed in the lowest panel of Figure S2, where in the lower frequency region the discrimination of distinct RF peaks would be no longer straightforward.

On the basis of these simulations, the cavity region was chosen to be distributed over 27 central holes, whose sensing area is about 12 $\mu m^2$.

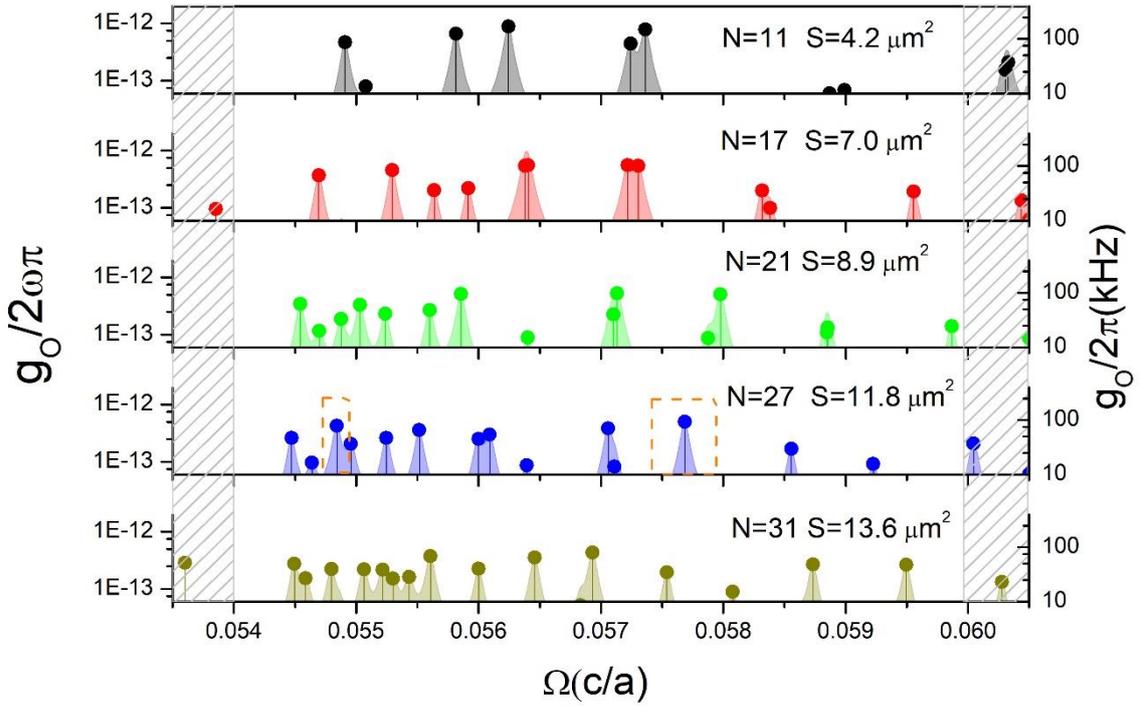

**Figure S2.** Normalized optomechanical coupling rates for different values of the total number of cells building the cavity region. The right vertical axis represents $g_O/2\pi$ considering a geometry scaled so that the optical mode appears around 180 THz. The dashed regions define the allowed region for cavity modes, which lies between the band edge values for $\Gamma=0.9$ (left) and $\Gamma=1$ (right). The highlighted peaks are associated to the couples of optical and mechanical modes considered for the illustrations of Fig. S4.

In Figure S3 we show the computed optical modes as a function of N while keeping fixed $\Gamma=0.9$. The OMC has been scaled so that the modes appear above 180 THz, thus falling in the spectral range covered by the tunable laser. It is also observed that the confined modes appear in the region between the bad edge values for Γ=0.9 and Γ =1.

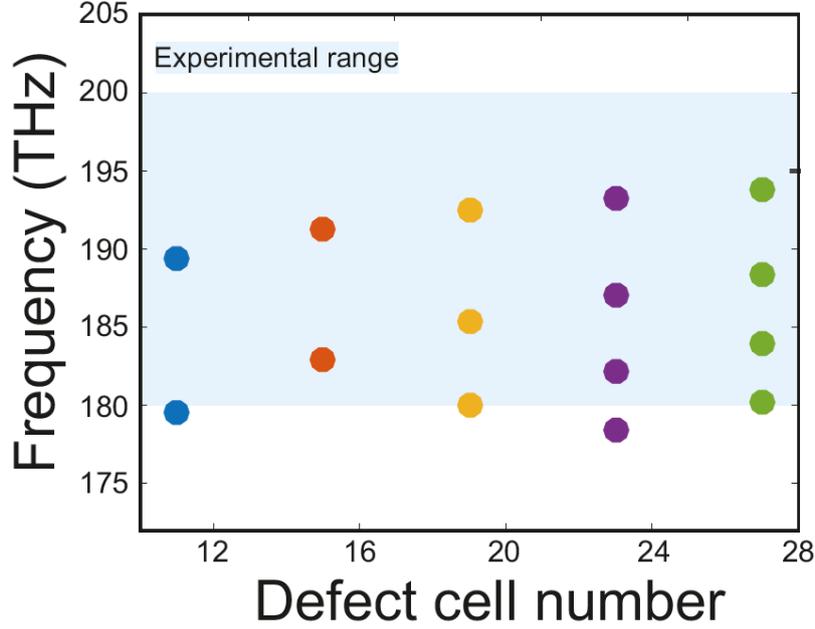

**Figure S3.** Optical modes as a function of the number of cells building the cavity region. The coloured area illustrates the spectral range covered by the tunable laser.

To illustrate better the optomechanical properties of this particular OMCs in Figure S4 we represent the different spatial contributions relevant for the calculation of $g_O$ for the particular realization having $N$=27 and Γ=0.9. In particular, we have focused on the optical mode in Figures S4a and S4e, and the mechanical modes in Figures S4b and S4f, which correspond to those highlighted in Figure S2 with dashed orange boxes. The optical mode is one appearing close to the band edge and thus extends over the whole cavity region. As a consequence, $g_O/2\pi$ takes values (greater than 10 kHz considering a geometry scaled so that the optical mode appears at 180 THz) for the whole family of pinch cavity modes, as also illustrated in Figure S2. The mechanical mode appearing in the high (low) frequency region $\Omega(a/c)$= 0.577 (0.548) is localized away from (close to) the center. In addition, both involve the oscillation of just a couple of adjacent cells, which is in agreement with what discussed above. This is also evidenced in Figures S4c and S4g, and Figures S4d and S4h, where we illustrate the PE volume density (the integrand of Eq. S2) and the MI surface density (the integrand of Eq. S1), respectively. Indeed, the volume contributing significantly to $g_O$ is reduced to the coloured regions. It is also worth mentioning that, in the pinch modes of the current study, $g_O$ is dominated by the MI contribution given that the PE contribution is an order of magnitude smaller.

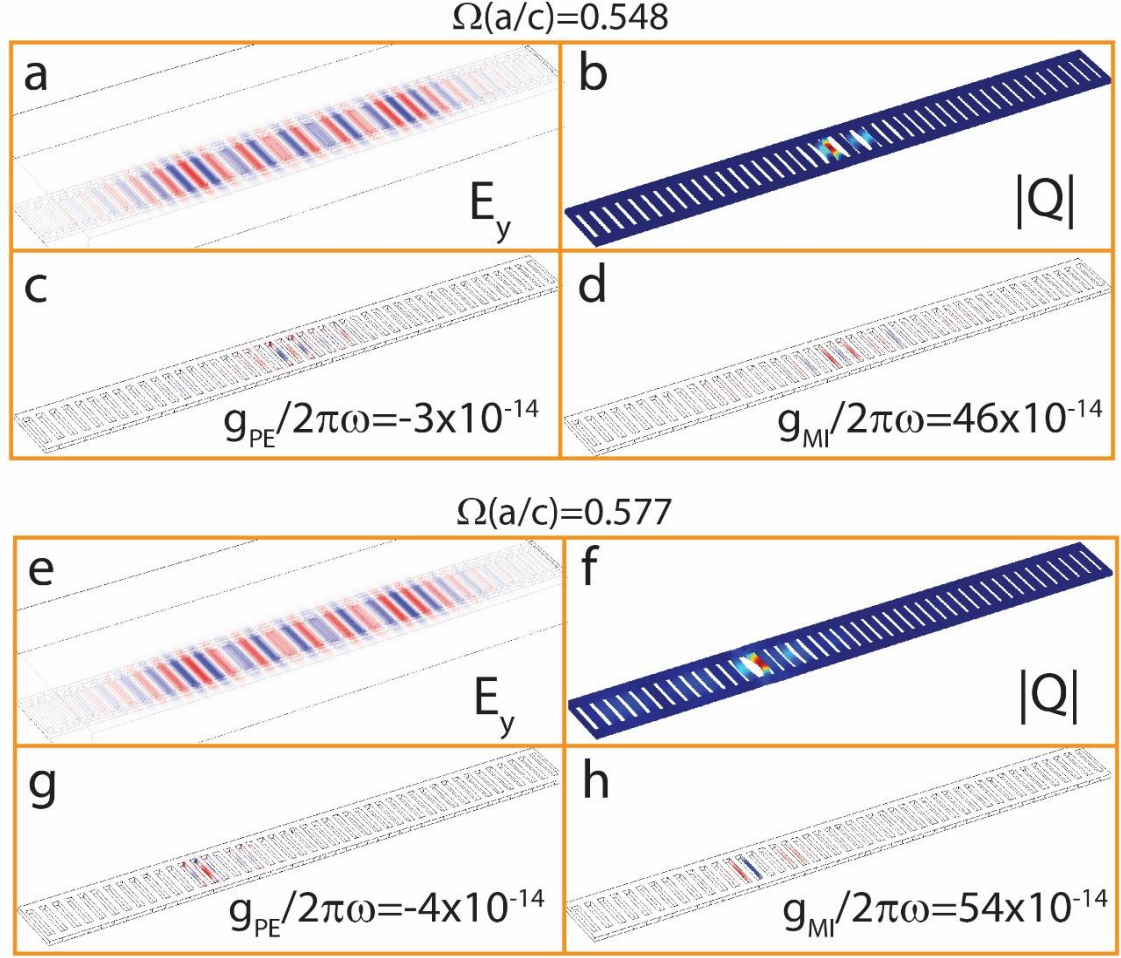

**Figure S4. Optical and mechanical modes of the Optomechanical Crystal cavity with N=27 and G=0.9 cells considering mechanical modes at $\Omega(a/c)=0.548$ and $0.577$ (top and bottom panels, respectively).** a) and e) Normalized optical $E_y$ field of the optical mode supported by the OM crystal used for the calculations reported in Fig. S2. b) and f) Normalized mechanical displacement field $|Q|$ of the pinch mechanical mode. c) and g) Normalized volumetric density of the integrand of the integrand in Eq. S2, showing the contributions to $g_{PE}$. d) and h) Normalized surface density of the integrand in Eq. S1, showing the contributions to $g_{MI}$.

## S4. Fabrication of the Optomechanical Crystal

The structures were fabricated in Silicon-on-Insulator (SOI) samples with a top silicon layer thickness of 250 nm (resistivity $\rho \sim 1-10$ Ohm.cm, p-doping of $\sim 1\times 10^{15}$ cm$^{-3}$) and a buried oxide layer thickness of 2 μm. The OMC cavities fabrication process was based on electron beam direct writing on a coated 170 nm of CSAR resist layer. The electron beam exposure was optimized with an acceleration voltage of 10 KeV and an aperture size of 30 μm with a Raith150 tool. After developing, the resist patterns were transferred into the SOI samples by inductively coupled plasma reactive ion etching. Finally, the silicon dioxide under the membranes was removed by using a HF bath.

## S5. Influence of particle elastic properties on spectral position of affected pinch modes

We have performed FEM simulations using the geometry imported from the SEM micrograph showed in Figure 2 in which we have modified the Young modulus and density of the particle. We verified that the affected pinch mode does not change further its frequency or spatial distribution in a wide range of input values around those of bulk $SiO_2$ (left and central panels of Figure S5). We have also shifted down the particle position along the crossbar (right panel of Figure S5), verifying that the mode is only slightly altered until when the particle starts loosing contact with the crossbar. In that situation the crossbar is released and the original mode displayed by the OMC without considering the particle is recovered.

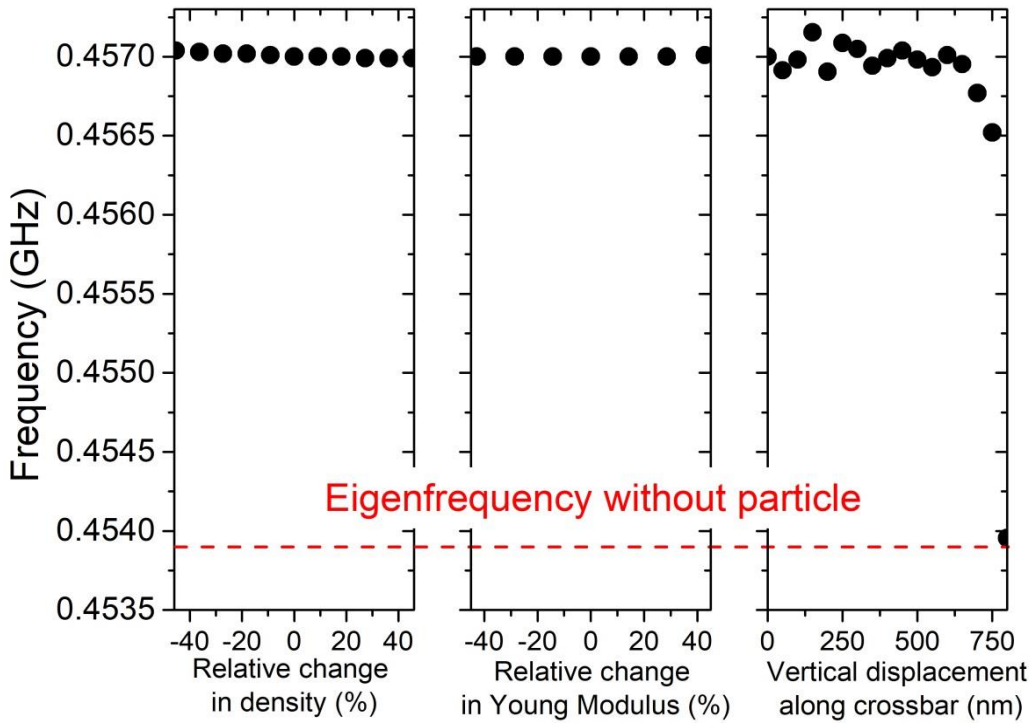

**Figure S5.** Dependence of eigenfrequency of the mechanical modes highlighted in Figure 2 of the main text with respect to relative changes in particle density (left panel) and Young modulus (central panel) and vertical displacement along the crossbar (right panel). The horizontal dashed line represents the eigenfrequency of the original mode without considering the particle on top of the OMC.

## S6. Pinch like mechanical modes of the fabricated Optomechanical Crystal involving the oscillation of the submicron particle

The presence of the submicron particles modifies the pinch mechanical mode spectrum as reported in Figures 2 and 3 of the main text. In addition, there are other worth mentioning modifications of the mechanical spectra. The first one is the onset of pinch-like modes that involve the collective oscillation of the particle and the bar in contact with it. These latter modes appear at significantly lower frequencies than those plotted in Figure 2, given that their effective mass is about a factor of two larger (Figure S6).

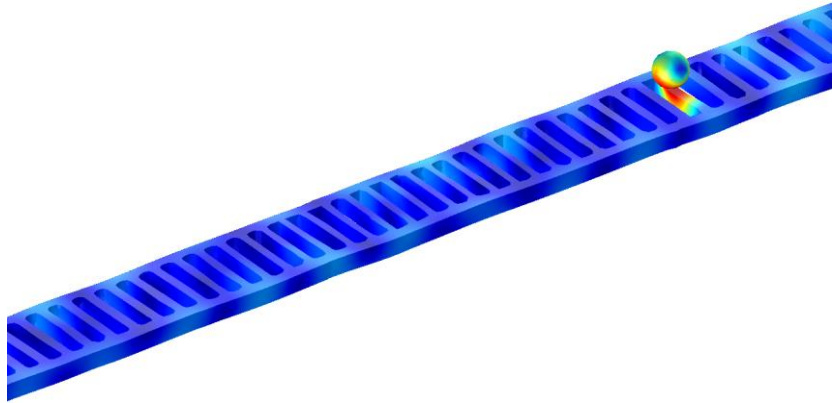

**Figure S6.** Pinch-like mode involving the collective oscillation of the particle and the bar in contact with it. This mode appears at significantly lower frequencies (0.35 GHz) than those plotted in Figure 2 of the main text, given that their effective mass is about a factor of two larger (about 0.2 pg).

Mechanical modes involving solely the deformation of the silica particle ($d$=500 nm) start appearing at much higher frequencies (few GHz, see Figure S8), in consistence to what obtained on an isolated particle of the same size.

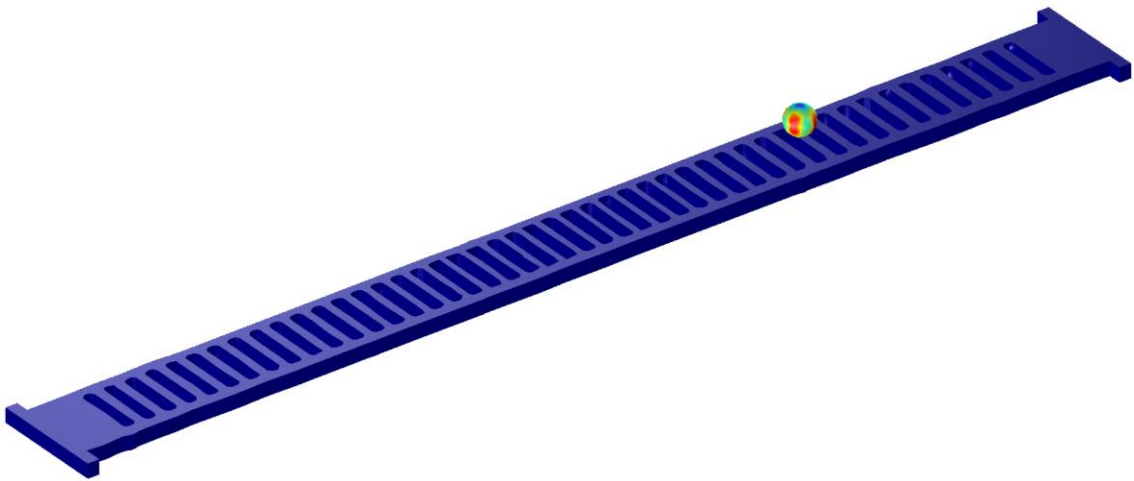

**Figure S7**. Mechanical mode involving solely the oscillation of the spherical particle. The calculated frequency of this particular mode is about 5.8 GHz.

We have also modified the Young modulus of the particle and its position along the crossbar and verified that the modified mode does not change further its frequency or spatial distribution. The previous statement holds unless there are mechanical eigenfrequencies of the isolated particle that are similar to those of the original pinch modes, in which case the modes hybridize similarly to what experimentally demonstrated in Ref.[4]. In Figure S8 we show that this is in fact the case if the elastic properties and the size of the spherical particle are modified to bring down its oscillation frequency to the frequency range of the pinch modes, i.e., to about 0.48 GHz.

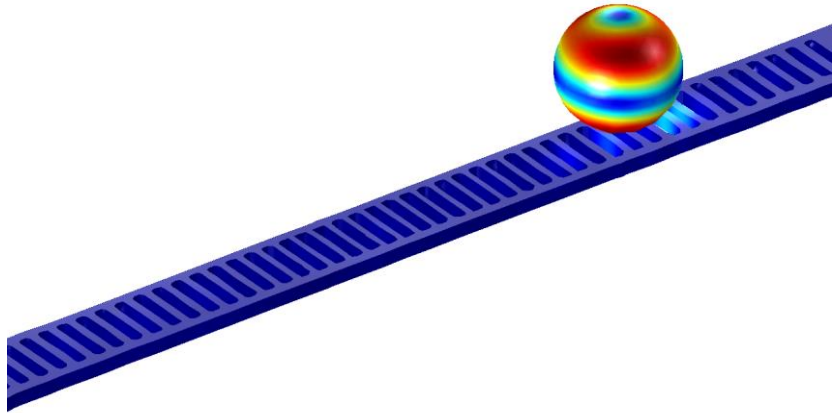

**Figure S8.** Pinch-like mode in which the natural mode of the particle hybridizes with a pinch mode of similar frequency. The elastic properties and the size of the spherical particle have been modified to bring down its oscillation frequency to about 0.48 GHz, but its positioning is the same as in the case considered in Fig. 2 and Fig. S6.

### S7. Eigenvibrations of the Silica Nanoparticles

The eigenmode vibration spectrum of the present silica nanoparticles (diameter $d = 495 \pm 16$ nm recorded by spontaneous Brillouin light spectroscopy (BLS) [5] is shown in Figure S9 (anti-stokes side). The lowest frequency quadrupolar (1,2) mode appears at 3.67GHz is much higher than the OMC resonance frequency. Based on the $f(1,2) = 0.85 c_t/d$, the transverse speed of sound is $c_t = 2140$ m/s [5].

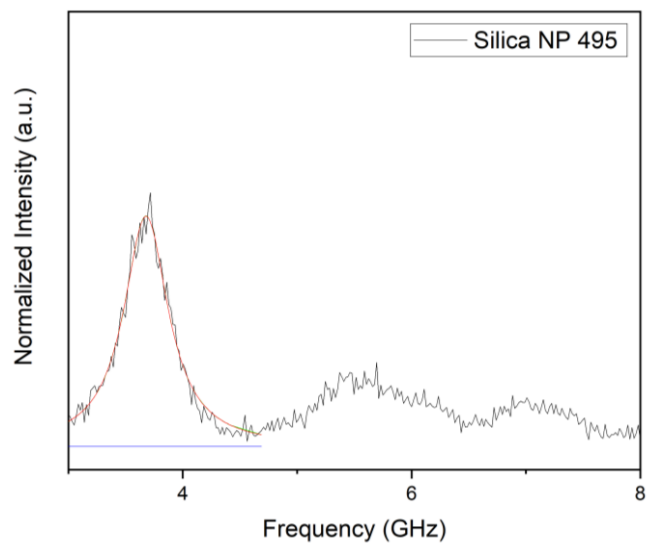

**Figure S9.** BLS eigenmode spectrum (anti-Stokes side) of spherical silica nanoparticles with diameter $d = 495 \pm 16$ nm.